\documentclass[
reprint,
nofootinbib,
amsmath,amssymb,amsbsy,
prd,
twocolumn,
superscriptaddress,
longbibliography,
preprintnumbers]{revtex4-2}

\usepackage[T1]{fontenc}
\usepackage{gfsartemisia-euler}
\usepackage{enumitem}
\usepackage{pifont}
\usepackage{amsmath}

\usepackage[mathscr]{euscript}

\makeatletter
\newcommand{\subalign}[1]{%
  \vcenter{%
    \Let@ \restore@math@cr \default@tag
    \baselineskip\fontdimen10 \scriptfont\tw@
    \advance\baselineskip\fontdimen12 \scriptfont\tw@
    \lineskip\thr@@\fontdimen8 \scriptfont\thr@@
    \lineskiplimit\lineskip
    \ialign{\hfil$\m@th\scriptstyle##$&$\m@th\scriptstyle{}##$\hfil\crcr
      #1\crcr
    }%
  }%
}
\makeatother

 \usepackage{lmodern}
\usepackage{graphicx}
\usepackage{hyperref}
\usepackage{amsfonts}
\usepackage[named]{xcolor}
\usepackage{tocvsec2}
\usepackage{slashed}
\usepackage{cancel}
\usepackage{comment}

\newcommand\beq{\begin{eqnarray}}
\newcommand\eeq{\end{eqnarray}}

\newcommand{\half}{\frac{1}{2}}

\newcommand\eqn[1]{\label{eq:#1}} 
 
\newcommand\Eq[1]{Eq.~(\ref{eq:#1})}

\newcommand{\bfx}{{\mathbf x}}

\newcommand{\CD}{{\cal D}}

\newcommand{\CJ}{{\cal J}}

\newcommand{\mybar}[1]{\kern 0.6pt\overline{\kern -0.6pt#1\kern -0.6pt}\kern 0.6pt}

\def\half{\tfrac{1}{2}}

\def\U\Omega{U(1)_{\Omega}}

\hypersetup{
  colorlinks=true,   
  linkcolor=cyan,    
  citecolor=blue,    
  filecolor=magenta, 
  urlcolor=blue      
}
\usepackage[section]{placeins}

\begin{document}

\title{Chiral Gauge Theory at the Boundary between Topological Phases}

\author{David B. Kaplan}
\email{dbkaplan@uw.edu}
\affiliation{Institute for Nuclear Theory, Box 351550, Seattle, Washington 98195-1550}
\preprint{INT-PUB-23-045}

\begin{abstract}
I demonstrate how chiral fermions with an exact gauge symmetry  can appear on the $d$-dimensional boundary of a finite volume $(d+1)$-dimensional manifold, without any light mirror partners. The condition for the $d$-dimensional boundary theory to be local is that gauge anomalies cancel and that the volume be large.  This can likely be achieved on a lattice and provides a new paradigm for the lattice regularization of chiral gauge theories.

\end{abstract}

\maketitle

If understanding a theory means that one can reliably compute consequences of it, then we do not understand the standard model---despite tremendous agreement between experimental data and calculations to low orders in perturbation theory in the weak interactions.  That is because there does not exist a nonperturbative regulator for chiral gauge theories, such as the standard model,  due to the tension in quantum field theory between chiral symmetry and the need to tame UV divergences.  Regulating a quantum field theory requires introduction of a UV mass scale, which typically breaks chiral symmetry. This tension has physical consequences in the form of anomalies, but it also places an obstacle to defining chiral gauge theories,  where violating chiral symmetry entails breaking gauge invariance.  

The tension between chiral symmetry and regularization applies equally to continuum and lattice approaches \cite{karsten1981lattice,Nielsen:1980rz}.  Domain wall fermions were introduced to solve the problem of realizing chiral symmetry on the lattice by exploiting the fact that  in infinite volume, massless fermion modes must exist on the spatial boundary between two  topologically distinct phases \cite{Kaplan:1992bt,Jansen:1992tw,Golterman:1992ub}.  Such phases can be realized by Wilson fermions, depending on the ratio  of the fermion mass to the Wilson coupling (the coefficient of $\bar\psi \partial^2\psi$) \cite{Golterman:1992ub}.  A  natural consequence of this  lattice construction is being able to correctly account for the anomalous divergences of global chiral currents and explicitly observe them \cite{jansen1992chiral}.
For actual computations on a finite lattice, the topological phase boundaries  can be approximately realized as the two edges of a finite five-dimensional lattice    \cite{shamir1993chiral}.   Our 4D world then consists of the two disconnected boundaries of a 5D slab with chiral modes of opposite chirality bound to the two surfaces, with 4D gauge fields independent of the extra dimension.  This gives a light  Dirac fermion with a  mass that vanishes exponentially fast with the extent of the extra dimension.  In the limit of an infinite extra dimension, chiral symmetry is restored (up to anomalies), and the massless chiral modes can be described by the overlap operator on a finite four-dimensional lattice, without reference to the gapped modes in the bulk \cite{Neuberger:1997fp,neuberger1998vectorlike,Narayanan:1993sk,Narayanan:1993zzh,Narayanan:1993ss}. In turn, the overlap operator solves the Ginsparg-Wilson equation \cite{Ginsparg:1981bj} which implies the existence of exact chiral ``L\"uscher'' symmetries of the lattice action which give rise  to the appropriate anomalous transformation of the lattice integration measure \cite{Luscher:1998pqa}.  

The domain wall and  overlap constructions have led to a complete solution to the problem of constructing vectorlike gauge theories with global chiral symmetries, but do not directly lead to a regulator for lattice chiral gauge theory.  This is because for every Weyl fermion on one boundary, there is a mirror fermion with opposite chirality on the other boundary, so the overall theory is vectorlike.  There have been numerous attempts to find a lattice regularization of a chiral gauge theory, using domain wall or overlap fermions as a starting point, or by other approaches \footnote{One approach has been to gauge the  L\"uscher symmetries, which has only been proved  possible for Abelian gauge symmetries \cite{luscher1999abelian,luscher2001chiral,Kadoh:2007wz}.  The ``symmetric mass generation'' approach has been to gauge the vector symmetry of domain wall fermions, and then attempt to gap the mirror fermion modes on one of the boundaries via strong localized four-fermion interactions  with recent work found in Ref.~\cite{PhysRevB.85.085103,PhysRevB.90.245120,You:2014vea,Wang:2013yta,Kikukawa:2017ngf,Wang:2018ugf,Catterall:2020fep,Wang:2022ucy,catterall2023lattice}; this approach has recently been criticized in Ref.~\cite{golterman2023propagator}.   An Abelian chiral gauge theory has recently been constructed on the lattice in 2D  using bosonization techniques  \cite{berkowitz2023exact}. Finally,  in perturbation theory it is possible to fine tune a chiral theory with broken gauge symmetry to the appropriate chiral continuum limit \cite{golterman2004s}, suggesting that maintaining gauge invariance on the lattice might not be necessary.}.  

Here I present an alternative, that a $d$-dimensional chiral gauge theory can be realized as  the boundary theory of a quantum field theory constructed on a finite $(d+1)$-dimensional manifold that has only a single, connected boundary.  With the manifold having only a single boundary, there are no mirror fermions.  I conclude that it is precisely the gauge anomaly cancellation condition that allows this system to behave as a local $d$-dimensional chiral gauge theory in the IR, and that there is no obvious obstruction to simulating such a system nonperturbatively on a lattice---although there remain unanswered questions (see Ref.~\cite{aoki2023curved} for related work).

Consider  a free massive Dirac fermion on the manifold $Y=M^{d-1} \times {\mathbb R}^2$ with Euclidian signature.  The $M^{d-1}$ manifold is described by the $d-1$ coordinates $\bfx_\perp$, while the ${\mathbb R}^2$ submanifold is described by Cartesian coordinates $\{x,y\}$ or polar coordinates $\{r,\theta\}$. The fermion mass is taken to equal $m$ for $r<R$ and $-M$ for $r>R$, with both $m$ and $M$ real and positive. I will eventually take $M\to\infty$ which will allow ignoring the region $r>R$, in which case $Y =M^{d-1} \times\mybar D_R $, where $\mybar D_R$ is the closed disk of radius $R$, and the boundary of $Y$ is $ M^{d-1} \times S^1$, which will serve as our $d$-dimensional spacetime.  I  take $d$ to be even and the fermions to be Dirac, but the analysis can be generalized to include Majorana fermions and edge states in odd spacetime dimensions, such as recently discussed in Ref.~\cite{clancy2023generalized}. 

The fermion action may be written as
\beq
S = \int d\bfx_\perp \int r\,dr\,d\theta\, \mybar \psi\left(\slashed{\partial}_\perp +\CD \right)\psi\ ,
\eqn{act}
\eeq
where $\slashed{\partial}_\perp= \vec\gamma_\perp \vec\partial_\perp$ is the Dirac operator on   $M^{d-1}$ and
\beq
\CD &=& \gamma_x \partial_x + \gamma_y\partial_y +m(r)
\cr &&\cr
&=&
\gamma_r \left(\partial_r+ \frac{1}{2r}\right) + \frac{i}{r}\gamma_\theta \CJ  + m(r)
,
\eqn{CDdef}\eeq
where
\beq
 \gamma_r = \cos\theta\,\gamma_x + \sin\theta\,\gamma_y ,\ \ 
  \gamma_\theta = -\sin\theta\,\gamma_x + \cos\theta\,\gamma_y ,
\eeq
  and $\CJ$ is the angular momentum operator
\beq
\CJ = -i \partial_\theta + \half\Sigma\ ,\qquad \Sigma = - \frac{i}{2}\left[\gamma_x,\gamma_y\right]\ .
\eqn{Jeq}\eeq
  Since $\CD $ is not Hermitian it is convenient to expand $\psi$ and $\bar\psi$ in the functions $f_n$ and $b_n$ respectively, which satisfy
\beq
\CD f_n = \mu_n b_n\ ,\qquad \mybar\CD b_n =\mu_n^* f_n\ ,
\eqn{bfeq}
\eeq
where 
\beq
\mybar\CD=
\frac{1}{r} \CD^\dagger r&=&
- \gamma_r \left(\partial_r+ \frac{1}{2r}\right) - \frac{i}{r}\gamma_\theta \CJ  + m(r)\cr 
&=&
\Sigma \CD \Sigma 
\eqn{CDbardef}\eeq
is the adjoint of $\CD$ with respect to the integration measure in polar coordinates.  As $f,b$ are eigenstates of the self-adjoint operators $\mybar\CD \CD$ and $\CD\mybar\CD$ respectively, they each can be taken to be a complete orthonormal basis. 
The magnitude of $\mu_n$ may be found by solving the eigenvalue equation
\beq
\mybar \CD \CD f_n = |\mu_n|^2 f_n\ ,
\eeq
and the phase of $\mu_n$, can be conveniently fixed by choosing
\beq
b_n = \Sigma f_n\ .
\eqn{bsf}\eeq
Only solutions with low lying eigenvalues $\vert \mu\vert < m$  correspond to boundary states.  

The cylindrical symmetry of the problem can be exploited by taking $f$ and $b$ to be eigenstates of the angular momentum operator $\CJ$  which commutes with both $\CD$ and $\mybar \CD$ and has eigenvalues $j=\pm\half,\pm\frac{3}{2},\ldots$.  Therefore a convenient basis to work in is one where the spin $\Sigma$ is diagonal, such as
\beq
\vec \gamma_\perp = \sigma_3\otimes \vec\Gamma\ ,\quad
\gamma_x = \sigma_1\otimes 1\ ,\quad
\gamma_y = \sigma_2\otimes 1\ ,
\eqn{basis1}\eeq
\beq
\Sigma = -\frac{i}{2}\left[\gamma_x,\gamma_y\right] = \sigma_3\otimes 1\ .
\eqn{basis2}\eeq
where $\vec \Gamma$ are the $2^{d/2-1}\times 2^{d/2-1}$ Dirac matrices in $(d-1)$ dimensions (for example,  $\vec\Gamma=1$   for $d=2$, and $\vec\Gamma = \vec\sigma$ for $d=4$).  In polar coordinates one has
\beq
 \gamma_r =\begin{pmatrix} 0 & e^{-i \theta} \\ e^{i \theta} & 0 \end{pmatrix}\ ,\qquad
  \gamma_\theta 
  =\begin{pmatrix} 0 & -ie^{-i \theta} \\ ie^{i \theta} & 0 \end{pmatrix}
\eqn{basispolar}
\eeq
while $\Sigma$ is unchanged. 

The fields $\psi$ and $\bar\psi$ can now be expanded  as
\beq
\psi_{\alpha i} &=& \sum_n f_{n,i}(r,\theta) \chi_{n,\alpha}(\bfx_\perp)\cr
\bar\psi_{\alpha i} &=& \sum_n \bar\chi_{n,\alpha}(\bfx_\perp) b^\dagger_{n,i}(r,\theta) \ ,
\eeq
where the spinor index $i=1,2$ is acted on by the first block in our direct product notation for the Dirac matrices,  the $\alpha=1,\ldots,2^{d/2-1}$ indices are acted on by the second block, and the  $\chi_{n,\alpha}(\bfx_\perp)$ are $2^{d/2-1}$-component spinors. 
Given the relations
\beq
\CD f_n = \mu_n b_n, \quad
b^\dagger_n \slashed{\partial}_\perp = f_n^\dagger\Sigma \slashed{\partial}_\perp = f_n^\dagger\, ( 1\otimes \vec\Gamma)\cdot\vec\partial_\perp \ ,
\eeq
 it follows that
the action in \Eq{act} can be then be rewritten as the sum of an infinite tower of fermions propagating on $M^{d-1}$:
 \beq
 S = \int d\bfx_\perp\, \sum_n \bar\chi_n\left(\vec\Gamma\cdot\vec\partial_\perp+\mu_n\right)\chi_n\ .
\eqn{Seff} \eeq

The unnormalized solutions to \Eq{bfeq} for the boundary modes on the disk ($r\le R$) in the $M\to\infty$ limit are given by
\beq
 f_j(r)  &=&\begin{pmatrix}
    e^{i(j-1/2)\theta}\frac{I_{|j-1/2|}( \kappa_j r)}{I_{|j-1/2|}( \kappa_j R)} \\[8 pt]  
    - e^{i(j+1/2)\theta}\frac{I_{|j+1/2|}( \kappa_j r)}{I_{|j+1/2|}( \kappa_j R)}\end{pmatrix}\  ,\cr &&\cr &&\cr
     b_j(r)  &=&  \begin{pmatrix}
    e^{i(j-1/2)\theta}\frac{I_{|j-1/2|}( \kappa_j r)}{I_{|j-1/2|}( \kappa_j R)} \\[8 pt]  
     e^{i(j+1/2)\theta}\frac{I_{|j+1/2|}( \kappa_j r)}{I_{|j+1/2|}( \kappa_j R)}\end{pmatrix}\ ,
 \eqn{bfeqsol}
 \eeq
where $I_\nu(z)$ is a modified Bessel function, 
$\kappa_j = \sqrt{m^2 - |\mu_j|^2}$, and  $\mu_j$ solves the implicit eigenvalue condition
\beq
\mu_j = m - \kappa_j\frac{I_{|j-1/2|}( \kappa_j R)}{I_{|j+1/2|}( \kappa_j R)}\ .
\eqn{musol}
\eeq
In the limit $M\to\infty$ one finds that the $f$ and $b$ solutions obey chiral boundary conditions at the edge of the disk,
\beq
\frac{1+\gamma_r}{2}f_j(R) =\frac{1-\gamma_r}{2}b_j(R) =0\ ,
\eeq
with $\gamma_r$ playing the role of $\gamma_5$.
in addition to the surface mode solutions there are less interesting bulk excitations  labeled by a radial excitation quantum number as well as $j$.

The eigenvalue equation \Eq{musol} can be 
 solved explicitly in an expansion in inverse powers of $m R$, with the result
 \beq
\mu_j &=& -\frac{j}{R}\left[1+\frac{1}{2 m R}+\frac{1}{2 m^2
   R^2}+\frac{3}{4 m^3 R^3}+\frac{3}{2 m^4 R^4} \right] 
   \cr&&+ \frac{j^3}{R}\left[\frac{1}{4 m^4 R^4}
   \right] + O\left((m R)^{-5}\right), 
  \eqn{musol2} \eeq
which is valid for either sign of $j$.  

To interpret the boundary mode action given in \Eq{Seff} with the above expression for $\mu_j$, compare with the Dirac operator in $d$ dimensions    in a chiral basis for the $\gamma$ matrices,
\beq
\vec \gamma_\perp &=& \sigma_1\otimes \vec\Gamma\ ,\ 
\gamma_\parallel = \sigma_2\otimes 1\ ,\ 
\gamma_\chi = \sigma_3\otimes 1\ ,
\eeq
\beq
\slashed{\partial} = \begin{pmatrix}
0 & \vec\Gamma\cdot\vec\partial_\perp -i \partial_\parallel \\
   \vec\Gamma\cdot\vec\partial_\perp +i \partial_\parallel  & 0
   \end{pmatrix}
   \eeq
   where $\vec\partial_\perp$ is the gradient in the $(d-1)$ dimensions and $\partial_\parallel = \partial/\partial x_d$. The upper right and lower left blocks of this matrix can be identified as the fermion operators for the left- and right-handed Weyl components of the Dirac fermion respectively.  If one were to compactify the $d$th dimension to a circle of radius $R$ and Fourier transform with respect to $x_d$, one would make the replacement $-i\partial_\parallel\to j/R$ in the above matrix, where $j$ takes integer values for periodic boundary conditions, and half integer values for antiperiodic. 
 Comparing this result with the fermion operator in \Eq{Seff}, along with the expansion of $\mu$ in \Eq{musol2}, one sees that the edge state can be identified with a right-handed Weyl fermion on a circle with antiperiodic boundary conditions, its linear momentum proportional to $j$. The opposite chirality would result if we reversed the signs of the masses $m$ and $M$.
The corrections in powers of $1/(m R)$ in \Eq{musol2} are due to the finite, $j$-dependent extent of the boundary state wave functions  into the bulk a distance $O(1/m)$.    To order $1/(m R)^3$  they are just renormalizing the value of $R$ that appears in the $j/R$ expression.  The $j^3$ contribution
at $O[(m R)^{-4}]$     corresponds to an irrelevant  three-derivative contribution to the kinetic term of the Weyl fermion, which does not violate chirality.  
While the $j^3$ term corresponds to an irrelevant operator, its appearance suggests that the dispersion relation could become nonanalytic in $j$ for $j\gtrsim mR$.  Indeed, that appears to be the case: a graphical solution of \Eq{musol} shows two eigenvalues merging and going into the complex plane for $j$ roughly equal to $m R$.

The result that an exactly chiral mode exists   on the $d$-dimensional boundary of a finite $(d+1)$-dimensional manifold  may seem counterintuitive.  If one were to elongate the disk, the system would look similar to the traditional wall/antiwall system which supports a right-handed edge state on one side, a left-handed one on the other, and an exponentially small but nonzero mass term from the overlap of their wave functions.  The reason why one does not find opposite chiralities on opposite sides of the disk is because while $\gamma_5$ is a constant matrix for the  wall/antiwall system, its analog for the disk is $\gamma_r$ in \Eq{basispolar} which is $\theta$-dependent,  changing sign from one side of the disk to the other, explaining how modes on opposite sides can have the same chirality.   The exponentially small interaction between the two modes on opposite sides of the finite disk can be seen by analytically continuing the eigenvalue equation \Eq{musol} to $j=0$ from either the $j\ge \half$ or $j\le-\half$ side and finding $\mu\sim \mp 2m\exp(-2 m R)$; in this case, however, such an interaction does not flip chirality, but instead  represents a nonlocality from the $d$-dimensional perspective which vanishes in the large $m R$ limit.  One should expect a corresponding nonanalyticity in the edge state dispersion relation for large $j\sim m R$.  It is unclear   whether the fact that the nonlocality is exponentially small in $R$ renders it innocuous.

I now turn to the question of gauging the theory.  The $(d+1)$-dimensional theory with $N$ copies of massive Dirac fermions  possesses a global $U(N)$ symmetry, any subgroup of which   can be gauged in a straightforward way with a well-defined integration measure for the path integral, once a regulator is included.   However, if one wants to describe a $d$-dimensional chiral gauge theory on the boundary, and not a theory of $d$-dimensional surface modes interacting with  $(d+1)$-dimensional gauge fields, one must find a way for the gauge fields in the bulk to be completely determined by their values on the surface, and not have independent bulk degrees of freedom.  Therefore one must define a gauge field $B_\mu$ over the whole disk  in such a way that it only depends on the gauge field's boundary value,
\beq
B_\mu(\bfx_\perp,r,\theta)\biggl\vert_{r=R} = A_\mu(\bfx_\perp,\theta)\ ,
\eqn{ABC}\eeq
where   $A_\mu$  is the field being integrated over in the path integral, subject to the usual measure $e^{-S_\text{YM}}$, where $S_\text{YM}$ is the $d$-dimensional Yang-Mills action.  

The role of gauge anomalies in defining a $d$-dimensional chiral gauge theory on the boundary is clarified using  anomaly in-flow arguments  \cite{Callan:1984sa}.  When integrating out the regulated bulk modes  in perturbation theory, the Chern-Simons operator involving $B_\mu$ is the only relevant operator that will be generated. With the $B_\mu$ fields being nonlocal functionals of the  $A_\mu$ gauge fields at the boundary, the existence of the Chern-Simons operator will in general preclude interpreting the theory of the edge states as being a local $d$-dimensional gauge theory. However, the exception is when the coefficient   of the Chern-Simons operator vanishes, which occurs precisely when the boundary theory is free of perturbative anomalies.  Therefore the conclusion is that when the $B_\mu$ fields are introduced and their boundary values $A_\mu$ are integrated over, the theory in the continuum will have a local $d$-dimensional description if and only if the perturbative gauge anomalies cancel.  
 This argument is extended to nonperturbative anomalies and made more  precise in Ref.~\cite{witten_anomaly_2020}.  That paper shows that the fermion determinant for the regulated system with a chiral fermion on the boundary takes the form $\sqrt{\det\slashed{D}} \exp(-i\pi \eta[B])$, where $\det\slashed{D}$ is the determinant of the massless Dirac operator, and $\eta$ is the $\eta$ invariant, the gauge invariant sum of signs of the eigenvalues of the bulk Dirac operator subject to generalized Atiyah-Singer-Patodi boundary conditions (which are defined so that the Dirac operator is self-adjoint) in the presence of the $B_\mu$ gauge field.
The $\eta$-invariant is perturbatively equivalent to the Chern-Simons operator, but contains additional information about nonperturbative anomalies.   When the edge theory is anomaly-free $\eta[B_\mu]$ is independent of the gauge field in the bulk and only depends on its boundary value, the physical gauge field $A_\mu$.

For practical applications one needs a concrete proposal for how to continue the gauge fields into the bulk. One possible definition, considered previously in Refs.~\cite{Grabowska_2016a,Grabowska_2016b} for related reasons, is to have $B_\mu$ be the solution to the Euclidian equations of motion subject to the  boundary condition \Eq{ABC}.
This is referred to as gradient flow, and has been widely used for unrelated applications  \cite{Luscher_2010,Luscher_2011}. If the boundary gauge fields $A_\mu$ are smooth compared to $m^{-1}$, this prescription ensures that the bulk fields $B_\mu$ will not change appreciably with respect to $r$ in the vicinity of the boundary, fulfilling the assumption of Ref.~\cite{witten_anomaly_2020}.  While gradient flow ensures that certain boundary gauge field configurations $A_\mu$ will lead to singularities in the bulk  gauge field $B_\mu$, these will be far from the boundary.

Finally, one must consider whether the theory can be realized on the lattice. Since the continuum chiral edge states arise at the boundary between two topological phases, and similar topological phases are known to exist for Wilson fermions on the lattice \cite{Jansen:1992tw,Golterman:1992ub}, there is no obvious obstacle for realizing the disk construction on the lattice.  One would apply the same open boundary conditions discussed in \cite{shamir1993chiral,luscher2013lattice}.  A recent lattice study of free Wilson fermions on a finite lattice with a single boundary corroborates the existence of a continuum Weyl fermion mode without any mirror partner \cite{Kaplan2023x}.
Such a spectrum would seem to violate the Nielsen-Ninomiya theorem and one of the assumptions going into that theorem must not apply. Since chiral symmetry and fermion number are synonymous for a Weyl fermion, and the underlying fermion number symmetry is exact for Wilson fermions, there is no violation of chiral symmetry, and hence it must be that one of the other assumptions is  being violated.
In fact, we have seen that the locality assumption underlying the Nielsen-Ninomiya theorem is violated due to the exponentially small interaction between fermions on opposite sides of the disk. For traditional domain wall fermions there were convincing effective field theory arguments for why global chiral symmetry became restored in the limit of infinite domain wall separation, but a precise  understanding of chirality in this system was not possible until the overlap description of the edge states was discovered and it was shown to obey the Ginsparg-Wilson equation \cite{Neuberger:1997fp}. 
 It is reasonable to wonder whether related reasoning can shed more light on how locality is realized in the model presented here.

Generically one can expect a sign problem that persists in the continuum limit when simulating a chiral gauge theory, but it is unknown how severe it will be. Perhaps one of the first applications of the idea should be the simulation of a Dirac edge state, where a sign problem does not exist in the continuum.  This would allow numerical exploration of potential nonlocality problems, as well as the expected dependence on solely the boundary values of the gauge fields.

This model gives reason for optimism that one can finally achieve a meaningful nonperturbative definition of chiral gauge theories such as the Standard Model. It can also be generalized to consider edge states in odd spacetime dimensions, as well as Majorana edge states. It is hoped that this formulation will allow chiral gauge theories to be simulated on a quantum computer some day --in order to overcome the sign problems and explore the rich phenomenology expected from them.

\ \

I  thank M. Golterman,   M. Savage, S. Sen,  Y. Shamir, and especially J. Kaidi and K. Yonekura for useful conversations and correspondence.
This research is supported in part by DOE Grant No. DE-FG02-00ER41132.

\bigskip

\bibliography{refs.bib}
\end{document}